\documentstyle[12pt,psfig]{article}
\def\simgt{\stackrel{>}{{}_\sim}}
\def\be{\begin{equation}}
\def\ee{\end{equation}}
\def\bear{\be\begin{array}}
\def\eear{\end{array}\ee}
\def\bea{\begin{eqnarray}}
\def\eea{\end{eqnarray}}

\begin{document}
\date{}



 

\def\talknumber{707} 

\title{Baryogenesis, Inflation and Superstrings}
\author{J.Alberto\,Casas\\
 Instituto de Estructura de la Materia, CSIC, Spain\\
(casas@cc.csic.es)}

\maketitle

\begin{abstract}
We study the conditions for successful Affleck--Dine baryogenesis 
in generic inflation and supergravity scenarios, finding powerful
restrictions on them. String-based SUGRA models are especially
interesting since they are surprisingly suitable for the implementation
not only of AD baryogenesis but also inflation itself, presenting a
nice solution to the $\eta$--problem.
\end{abstract}

\vskip-11.5cm
\rightline{}
\rightline{ IEM-FT-170/98}
\rightline{ January 1998}
\rightline{hep-ph/9802210}

\vspace{9.3cm}

\section{Introduction}

In the absence of better alternatives,
the Affleck and Dine (AD)  mechanism \cite{AD,DRT} is a very
attractive method for baryogenesis. It takes place
if in the early 
universe some scalar ``AD fields'', $\phi$, carrying baryon or 
lepton number ($B$ or $L$), get large initial
vacuum expectation values (VEV's), $\phi_{in}$. Then the equations
of motion of $\phi$ (together with the presence of some 
baryon-violating operator involving the $\phi$ fields) lead to
a net final baryon number \cite{AD}. The mechanism is quite natural
and efficient and does not need any particular tuning of parameters.
The key point is therefore how to generate the large initial VEV's,
$\phi_{in}$. 
These have been commonly associated in the
literature with the existence of (approximately) flat directions
involving $\phi$. Then quantum fluctuations during inflation 
may yield a large $\phi_{in}$. 

However, during inflation
SUSY is necessarily spontaneously broken since the scalar potential
$V$ gets a VEV, $\langle V \rangle= 3 H^2 M_{P}^2$, and effective 
SUSY soft breaking terms, in particular effective soft 
masses of the order of the Hubble constant $H$
are generated, which clearly spoils any flat direction \cite{DRT}.
Nevertheless, large initial VEV's are still possible if these
effective masses squared are {\em negative}, something that is 
possible in a generic SUGRA theory \cite{DRT}. The negative mass
destabilizes the potential at the origin and, for large values of
$\phi$, the potential is lifted, e.g. by F--terms coming from 
$\sim \phi^n$ terms in the superpotential $W$. Thus, a temporal 
minimum is generated. The $\phi$ field 
evolves rapidly towards the temporal minimum during this period.
For the subsequent evolution to be succesful for baryogenesis,
it is neccesary that the operators that lift the
potential are non-renormalizable (i.e. $n\ge 4$), 
which is perfectly possible \cite{DRT}.

To summarize, succesful AD baryogenesis requires that the $\phi$ 
effective potential {\em during inflation} contains
\begin{enumerate}

\item negative effective mass terms, $m^2_\phi\le 0$,

\item non-renormalizable terms to lift ``flat directions'' 
of the potential.

\end{enumerate}
%

\section{D--Inflation}

The possibility of D--inflation \cite{CM,Stew,BD} 
(i.e. inflation triggerred by a 
non-vanishing D--term) in SUGRA scenarios is very attractive
since, as has been often claimed in the literature, F--inflation seems
to lead naturally to too large inflaton mass terms that disable 
the inflationary process (see end of sect.4).

In order to be concrete, it is convenient to
suppose
that inflation is mainly triggered by a single D--term (this does not
reduce the generality of the analysis). Then, a suitable
choice is to suppose that the relevant D--term is associated to one
``anomalous'' $U(1)$ \cite{DSW,CKM}, which takes the 
form
\be
V_D=\frac{1}{2}D^2=\frac{1}{2}g^2\left|
\xi+\sum_j q_j|z_j|^2 K_{j\bar j} \right|^2\;,
\label{FI}
\ee
where $g$ is the corresponding gauge coupling, $q_j$ are the charges of
all the chiral fields, $z_j$, under the anomalous $U(1)$ and 
$K_{j\bar j}$ is the K{\"a}hler metric
(we are assuming here a basis for the $z_j$ fields 
 where the K{\"a}hler metric is diagonal). Finally, the 
constant $\xi$
is related to the apparent anomaly, $\xi=g^2M_P^2(\sum_j
q_j/192\pi^2)$. At low energy the D--term is cancelled
by the VEV's of some of the scalars entering Eq.(\ref{FI}), but
initially $\langle D\rangle$ may be different from zero, thus triggering
inflation.
Let us also notice that the scenario is quite insensitive to the
details of the K{\"a}hler potential $K$ (note in particular that
$(K_{j\bar j})^{1/2} z_j$ are simply the canonically normalized
chiral fields). 
%

Concerning AD baryogenesis, there are two main scenarios to 
consider, depending on $q_\phi\neq 0$ or $q_\phi = 0$.
In the first case, it is clear that the D--term induces an effective
mass term for $\phi$, which is {\em negative} provided
\be
{\rm sign}(\xi)\  {\rm sign}(q_\phi) =-1, 
\label{signs}
\ee
then the
effective mass squared is {\em negative} and we expect $\langle\phi
\rangle_{in}\neq 0$.

Nevertheless, this cannot be the whole story, since in the absence of
additional $\phi$--dependent terms in $V$, $\langle\phi
\rangle_{in}$ would adjust itself to cancel the D--term, thus disabling
the inflationary process and breaking $B$ or $L$ at low energy.
Thus, we need extra contributions yielding $\langle D
\rangle_{in}\neq 0$, $\langle\phi\rangle_f= 0$.  These may come from 
{\em (a)} low-energy soft breaking terms, {\em (b)} F--terms,
{\em (c)} D--terms. 
We do not have space to review in detail the three possibilities
(the interested reader is referred to ref.\cite{CG}). For our purposes,
it is enough to say that no one of them works for the goal of 
AD baryogenesis. The reason is that either the extra contributions 
are too small (case {\em (a)}) or they lead to a potential which
is lifted by renormalizable terms (contradicting condition 2 of
Sect.1).

This leaves us just with the second scenario, namely, $q_\phi=0$.
Then, from (\ref{FI}), $m_\phi^2=0$ during inflation. So, 
there is a truly flat direction along $\phi$ and the AD mechanism
can be implemented in the old-fashioned way after all!
This argument is only exact at tree level. Strictly speaking, there
are small contributions to $m_\phi$  coming from higher loop corrections
\cite{Dv} and the expected $O($TeV) low-energy supersymmetry breaking 
effects. In any case, $\phi$ will acquire a large VEV during inflation
due to quantum fluctuations if the correlation length for de Sitter 
fluctuations, $l_{coh} \simeq H^{-1}$exp(3$H^2/ 2 m_\phi^2)$
\cite{Bu}, is large compared to the horizon size. This translates
into the condition $H^2/m_\phi^2 \simgt 40$ \cite{DRT},
which is easily fulfilled in this context. So, the  $q_\phi=0$
scenario is really selected for AD baryogenesis, leading to 
very interesting physics

\section{F--Inflation}

F--Inflation occurs when inflation is triggerred by a 
non-vanishing F--term of the appropriate size.
Concerning the implementation of the AD mechanism in F--inflationary 
scenarios, the main question (see Sect. 1) is whether it is possible 
to get an effective mass squared $m_\phi^2<0$ 
or $m_\phi^2\simeq 0$ for the
AD field, $\phi$, during inflation.
To answer this question, we need to examine the F--part of the
effective potential $V$, in a SUGRA theory
\be
V= e^G\left( G_{\bar j} K^{\bar j l}G_{l}-3\right)
= F^{\bar l} K_{j \bar l}F^j-3e^G~.
\label{VFD}
\ee
Here $G=K+\log|W|^2$ where $W$ is the superpotential, $K^{\bar l j}$
is the inverse of the K{\"a}hler metric and 
$F^j=e^{G/2}K^{j\bar k}G_{\bar k}$ are the
corresponding auxiliary fields.
During inflation $\langle V \rangle_{in} = V_0 \simeq H^2 M_P^2$,
which implies that some $F$ fields are different form zero, thus
breaking SUSY. The effective gravitino mass squared 
during the inflationary 
epoch is given by  $m_{3/2}^2=e^{G}= e^{K}|W|^2$ in $M_P$
units. 
The SUSY breakdown
induces soft terms for all the scalars, in particular for $\phi$.
More precisely, the value of the effective mass squared, $m_\phi^2$, is
intimately related to the form of $K$. 
It is convenient to parametrize $K$ as 
\be
K = K_0(I) + K_{\phi\bar\phi} |\phi|^2 + \cdots\;,
\label{K}
\ee
where $I$ represents generically the inflaton or inflatons.
Plugging (\ref{K}) in (\ref{VFD}) it is straightforward to 
see \cite{CG}, and this is an important result, that if
there is no mixing between $\phi$ and $I$ in the quadratic term of $K$,
 i.e. if
 $K_{\phi\bar\phi}\neq K_{\phi\bar\phi}(I)$, then
the effective mass squared for the canonically normalized field, 
$(K_{\phi\bar\phi})^{1/2}\phi$, is
$m_\phi^2=m_{3/2}^2 + V_0/M_P^2$.
Hence $m_\phi^2$ is of $O(H^2)$ and {\em positive} and, therefore,  
the AD mechanism cannot be implemented. This excludes, for instance,
minimal SUGRA.
 
A successful implementation of the AD mechanism thus requires a mixing 
in the quadratic term of $K$ of the inflaton $I$
 and AD fields $\phi$, i.e.
$K_{\phi\bar\phi} = K_{\phi\bar\phi}(I)$ in Eq.(\ref{K}).
This mixing should be remarkably strong and even so the possibility
of a negative effective mass term is not guaranteed. 
For example, if we
consider the following simple scenario
$K = K_0(I) + |\phi|^2 + a|I|^2 |\phi|^2$,
where $a$ is some unspecified coupling, it is possible to see
after some algebra \cite{CG} that  for $|I|^2 > 3/4$ (in Planck units) 
there is no value of $a$ for which $m_\phi^2\le 0$. For smaller
values of $I$, negative masses squared are possible if $a> 1/3$. 
%

Finally, let us note that the possibility of a very small mass 
$m_\phi^2\sim 0$ (also welcome for a successful inflation itself, 
as we will discuss shortly)
does not seem natural at first sight, since it would imply
some conspiracy between the various contributions to $m_\phi^2$
coming from (\ref{VFD}).
However, the study of the SUGRA
scenarios coming from strings provides beautiful surprises in this
sense, as we are about to see.

\section{String Scenarios}

The best motivated SUGRA scenarios are those coming
from string theories. The corresponding K{\"a}hler potential, $K$,
is greatly constrained and, therefore, the implementation
of AD baryogenesis for the F-inflation framework is not
trivial at all. 
%
%
In order to be concrete we will consider orbifold constructions,
where the (tree--level) K{\"a}hler potential is given by \cite{DKL}
\be
K =  -\log(S+\bar S)-3\log(T+\bar T)+\sum_j(T+\bar T)^{n_j}|z_j|^2~.
\label{Kor}
\ee
Here $S$ is the dilaton and $T$ denotes generically the moduli 
fields, $z_j$ are the chiral
fields and $n_j$ the corresponding modular weights. The latter depend
on the type of orbifold considered and the twisted sector to which
the field belongs. The possible values of $n_j$  are 
$n_j= -1,-2,-3,-4,-5$. The discrete character of $n_j$ will play
a relevant role later on.

Since a strong mixing between the inflaton
and the AD field $\phi$ in the quadratic term 
 ($\propto |\phi|^2$) of $K$ is required (see Sect. 3), our first 
conclusion is that $T$ is the natural inflaton candidate
 in string theories. S--dominated inflation cannot work, and
this is a completely general result since the (tree-level)
S--dependence of K is universal in string theories.
We should recall, however, that a strong mixing in $K$ is a necessary
but not sufficient condition for $m_\phi^2\le 0$. We must then examine
the precise value of $m_\phi^2$ in the presence of a non-vanishing
cosmological constant $\langle V \rangle_{in}=V_0>0$. Restricting
ourselves to the moduli-dominated case, $m_\phi^2$ 
is given by \cite{BIM}
\be
m_\phi^2 = m_{3/2}^2\left\{(3+n_\phi)C^2-2\right\}~,
\label{mphi4}
\ee
where $m_{3/2}^2=e^{K}|W|^2$, 
$C^2= 1+ [V_0 /(3M_P^2m_{3/2}^2)]$ and $n_\phi$
is the modular weight of the AD field $\phi$. Since $C^2>1$,
it is clear that $n_\phi\le -3$ is a sufficient condition
to get $m_\phi^2\le 0$. States with $n_\phi\le -3$ occur in all
the orbifold  constructions, so $m_\phi^2\le 0$ is
perfectly natural, something that was not trivial at all
{\em a priori}. Actually, 
 if $C^2\le 2$, then $m_\phi^2\le 0$ whenever
$n_\phi\le -2$, which is a very common case.

There is a particularly interesting limit of eq.(\ref{mphi4}) 
that could well be realized in practice.
Namely, since 
$V_0= K_{T \bar T}|F^T|^2-3m_{3/2}^2=3H^2 M_P^2$, it may
perfectly happen that $K_{T \bar T}|F^T|^2\gg m_{3/2}^2$, and 
thus $C^2\gg 1$ (see definition of $C^2$ after eq.(\ref{mphi4})).
Then, from eq.(\ref{mphi4})
\be
m_\phi^2\simeq H^2(3+n_\phi). 
\label{msmall}
\ee
Hence for $n_\phi=-3$ we get $m_\phi^2\simeq 0$.
So, we see that the possibility of a very small mass
$m_\phi^2\simeq 0$  can occur in 
F--inflation, as it was the case in D--inflation. Notice that
there is no fine-tuning here, since $n_\phi$ is a discrete number
which can only take the values $n_\phi=-1,-2,-3,-4,-5$.

What is even more important:
this is also good news for F--inflation itself. As has been
pointed out in the literature, F--inflation
has the problem that if the inflaton mass is $O(H)$, as expected
at first sight during inflation, then the necessary slow rollover
is disabled (this is the so-called $\eta$--problem). 
We see here, however, that a hybrid--inflation
scenario \cite{Linde} in which $T$ is the field responsible for 
the large $V_0$ and a second field (any one with $n=-3$) is
responsible for the slow rollover is perfectly viable.
This is a nice surprise since F-inflation is very difficult to
implement in generic SUGRA theories, even with fine-tuning!



\def\MPL #1 #2 #3 {{\em Mod.~Phys.~Lett.}~{\bf#1}\ (#2) #3 }
\def\NPB #1 #2 #3 {{\em Nucl.~Phys.}~{\bf B#1}\ (#2) #3 }
\def\PLB #1 #2 #3 {{\em Phys.~Lett.}~{\bf B#1}\ (#2) #3 }
\def\PR  #1 #2 #3 {{\em Phys.~Rep.}~{\bf#1}\ (#2) #3 }
\def\PRD #1 #2 #3 {{\em Phys.~Rev.}~{\bf D#1}\ (#2) #3 }
\def\PRL #1 #2 #3 {{\em Phys.~Rev.~Lett.}~{\bf#1}\ (#2) #3 }
\def\PTP #1 #2 #3 {{\em Prog.~Theor.~Phys.}~{\bf#1}\ (#2) #3 }
\def\RMP #1 #2 #3 {{\em Rev.~Mod.~Phys.}~{\bf#1}\ (#2) #3 }
\def\ZPC #1 #2 #3 {{\em Z.~Phys.}~{\bf C#1}\ (#2) #3 }

\end{document}